\documentclass[10pt,journal,compsoc]{IEEEtran}
\usepackage{booktabs} 

\usepackage{amsmath}
\usepackage{mathtools}
\usepackage{listings}

\usepackage{url}

\usepackage{algorithmic}
\ifCLASSOPTIONcompsoc
  \usepackage[nocompress]{cite}
\else
  \usepackage{cite}
\fi

\begin{document}
\title{Loop Tiling in Large-Scale Stencil Codes at Run-time with OPS}

\author{Istv\'an~Z.~Reguly,~\IEEEmembership{Member,~IEEE,}
        Gihan~R.~Mudalige
        and~Michael~B.~Giles
\IEEEcompsocitemizethanks{
\IEEEcompsocthanksitem I.Z. Reguly is with PPCU ITK, Budapest, Hungary. Email: reguly.istvan@itk.ppke.hu
\IEEEcompsocthanksitem G.R. Mudalige is with Department of Computer Science, University of Warwick, UK. Email: g.mudalige@warwick.ac.uk
\IEEEcompsocthanksitem M.B. Giles is with the Maths Institute, University of Oxford, UK. Email: mike.giles@maths.ox.ac.uk
}
\thanks{Manuscript received XXX}}

\markboth{IEEE Transactions on Parallel and Distributed Systems,~Vol.~X, No.~X, X~201X}%
{Reguly \MakeLowercase{\textit{et al.}}: Loop Tiling in Large-Scale Stencil Codes at Run-time with OPS}

\IEEEtitleabstractindextext{%
\vspace{-10pt}\begin{abstract}
The key common bottleneck in most stencil codes is data movement, and prior research has shown that improving data locality through optimisations that schedule across loops do particularly well. However, in many large PDE applications it is not possible to apply such optimisations through compilers because there are many options, execution paths and data per grid point, many dependent on run-time parameters, and the code is distributed across different compilation units.  In this paper, we adapt the data locality improving optimisation called iteration space slicing for use in large OPS applications both in shared-memory and distributed-memory systems, relying on run-time analysis and delayed execution. We evaluate our approach on a number of applications, observing speedups of 2$\times$ on the Cloverleaf 2D/3D proxy application, which contain 83/141 loops respectively, $3.5\times$ on the linear solver TeaLeaf, and $1.7\times$ on the compressible Navier-Stokes solver OpenSBLI. We demonstrate strong and weak scalability up to 4608 cores of CINECA's Marconi supercomputer. We also evaluate our algorithms on Intel's Knights Landing, demonstrating maintained throughput as the problem size grows beyond 16GB, and we do scaling studies up to 8704 cores. The approach is generally applicable to any stencil DSL that provides per loop data access information.

\end{abstract}
\vspace{-5pt}
\begin{IEEEkeywords}
DSL, Tiling, Memory Locality, OPS \vspace{-5pt}
\end{IEEEkeywords}}

\maketitle
\IEEEdisplaynontitleabstractindextext
\IEEEpeerreviewmaketitle

\ifCLASSOPTIONcompsoc
\IEEEraisesectionheading{\section{Introduction}\label{sec:introduction}}
\else
\vspace{-15pt}
\section{Introduction}
\label{sec/intro}
\fi

\vspace{-5pt}
\IEEEPARstart{M}{odern} architectures now include ever-larger on-chip caches to help exploit spatial and temporal locality in memory accesses: latency and energy benefits of accessing data from cache can be up to 10x compared to accessing it with a load from off-chip memory. Unfortunately, most scientific simulations are structured in a way that limits locality: the code is structured as a sequence of computations, each streaming a number of data arrays from memory, performing a number of operations on each data element, then streaming the resulting arrays back to memory.

Improving memory locality is an area of intense research, and stencil codes have long been a target, given their regular memory access patterns and (mostly) affine loop structures. In stencil codes, we iterate through a 1/2/3 (or higher) dimensional grid, and perform computations given data on the current and adjacent grid points - the adjacency pattern is called the stencil. In a single loop nest (one sweep over the domain) there is already potential for data reuse, given the stencils used (e.g. along the contiguous dimension), which can be further improved using loop blocking \cite{blocking} - this is standard practice in modern compilers.

Loop fusion \cite{Darte} merges multiple subsequent loop nests into a single loop nest, making data re-use possible on a larger scale - across loop nests. It is easy to do when loop bounds of subsequent loop nests align, and data dependencies are trivial. There are many examples of loop fusion, demonstrating its importance \cite{6468479}. Loop fusion in the presence of non-trivial stencils (loop nest reading data generated by a previous loop nest with a multi-point stencil) is much more difficult because loops have to be shifted depending on the stencil pattern, leading to wavefront schemes.

There is a large body of research on the combination of fusion and loop schedule optimisations \cite{Pugh2000,Bertolacci,Muranushi}: techniques that extend loop blocking to work across subsequent loop nets, generally called \emph{tiling}. Tiling with iteration space slicing carries out dependency analysis similar to what is required for loop fusion, but instead of fusing the bodies of subsequent loops, it forms small blocks in each loop nest (fitting in the cache). Tiling achieves memory locality by executing the same set of blocks in subsequent loops, formed to satisfy data dependencies, then moves on to another set of blocks, etc. The key concept is that tiling achieves locality across a number of loop nests with non-trivial data dependencies across them. There is a well-established framework for loop scheduling transformations: the polyhedral framework.

Research into polyhedral compilers has laid a strong theoretical and practical foundation for cache blocking tiling, yet their use is limited by the fact that they apply compile-time optimisations. These compilers struggle with dynamic execution paths, where it is not known in what exact order loops follow one another, and they cannot manage analysis and code generation across multiple compilation units. Furthermore, many cannot handle branching that would lead to different access patterns within a single loop. Commonly used benchmarks come from SPEC OMP \cite{specomp} and PolyBench \cite{polybench}, but many of these issues do not come up, as these are constructed to be simple test cases. In summary, these compilers have primarily been shown to give excellent performance when a small number of loops repeat a large number of times in a predictable manner - which we do \emph{not} consider large-scale codes for the purposes of this paper.

The OPS (Oxford Parallel library for Structured meshes) DSL (Domain Specific Language) \cite{ops1,ops2} is essentially a C/C++/Fortran domain-specific API that uses source-to-source translation and various back-end libraries to automatically parallelise applications. Any code written using its API can utilise MPI, use multi-core CPUs with OpenMP, as well as GPUs with CUDA, OpenACC, or OpenCL. OPS is being used in a number of PDE applications \cite{ops2,jammy,jianping}, and indeed the common bottleneck in all of these applications is data movement. Unfortunately, the aforementioned challenges combined with the complexity of these applications prohibits the use of traditional stencil compilers, therefore we adopt an iteration space slicing algorithm \cite{Pugh2000}, referred to as tiling in the rest of the paper, that we apply at run-time using delayed execution of computations.

We choose the CloverLeaf 2D/3D code (part of the Mantevo suite) to demonstrate our results in detail, as it is a larger code that has been intensively studied by various research groups \cite{cloverleaf,ops2,6495848}; it is a proxy code for industrial hydrodynamics codes. It has 30 datasets (30 data values, or variables, per grid point), it consists of 83/141 different loops across 15 source files. During the simulation, a single time iteration consists of the execution of 150/600 loops, where often the same loop is executed on different datasets. Furthermore, some stencils are data-dependent, and there is considerable logic that determines the exact sequence of loops. To further demonstrate the utility of our approach, we evaluate performance on two more applications using OPS: the matrix-free sparse linear solver proxy code TeaLeaf \cite{tealeaf} (also part of the Mantevo suite), and the compressible Navier-Stokes solver OpenSBLI \cite{opensbli}.

In this paper, we present research into how, through a combination of delayed execution and dependency analysis, OPS is capable of addressing the aforementioned challenges, \emph{without modifications to the high-level OPS user code}. This is then evaluated through a series of benchmarks and analysed in detail. Specifically, we make the following contributions:
\vspace{-3pt}
\begin{enumerate}
	\item We introduce the delayed execution scheme in OPS and describe the dependency analysis algorithm that enables sliced tiling execution on a single block.
	\item We extend our algorithms to analyse dependencies and perform scheduling and communications in a distributed memory environment
	\item We validate and evaluate the proposed algorithm on a 2D Jacobi iteration example, comparing it to prior research (Pluto and Pochoir).
	\item We deploy the tiling algorithm on a number of larger-scale applications, such as the CloverLeaf 2D/3D hydrocode, TeaLeaf, and OpenSBLI. We explore relative and absolute performance metrics, including speedup, achieved bandwidth and computational throughput on Xeon server processors, scaling up to 4608 cores on CINECA's Marconi.
	\item We evaluate tiling on the Intel Knights Landing platform, scaling up to 128 nodes (or 8704 cores).
\end{enumerate}
\vspace{-3pt}
The rest of the paper is organised as follows: Section \ref{sec/related} discusses related work, Section \ref{sec/ops} summarises the design and implementation of OPS, Section \ref{sec/alg} presents the tiling algorithm integrated into OPS, Section \ref{sec/apps} introduces the applications we evaluate in this work and Section \ref{sec/bench} carries out the in-depth performance analysis. Section \ref{sec/mpi} evaluates strong and weak scaling on a CPU cluster, and  Finally Section \ref{sec/conc} draws conclusions.

\vspace{-10pt}
\section{Related work}\label{sec/related}
\vspace{-5pt}
Manipulating loop schedules to improve parallelism or data locality has long been studied and built into compilers \cite{Wolfe1986,Wolfe1989,blocking,Wolfe1982,polly}. The mathematics and techniques involved in such loop transformations have been described in the polyhedral framework \cite{polyhedra1,WILDE1993POLYLIB,GEIGL1997}, and since then, a tremendous amount of research has studied transformations of affine loop structures in this framework, and extended it to work on many non-affine cases as well.

Tiling by manually modifying code has been demonstrated on smaller codes  \cite{simd,Strzodka} where one or two loops repeat a large number of times (typically a time iteration); it is a particularly good example of utilising the large caches on CPUs, and they have been studied in detail.

There are a number of compilers focused on applying tiling to stencil computations such as Pochoir \cite{Tang:2011:PSC:1989493.1989508}, image processing workflows such as Polymage and Halide \cite{polymage,halide}, and more generally to computations covered by the polyhedral frameworks Pluto \cite{uday08cc,BASKARAN2009}, R-STREAM \cite{SCHWEITZ2006RSTREAM} - these have shown significant improvements in performance by exploiting data locality by manipulating loop schedules. There are examples of tiling in distributed memory systems as well: R-STREAM \cite{SCHWEITZ2006RSTREAM}, Pluto \cite{Bondhugula:2013:CAL:2503210.2503289}, Classen and Griebl \cite{1639500}.

The kinds of transformations applied are also wide-ranging, starting at the simplest skewed tiling methods across time iterations \cite{Strzodka,Wolfe1986}, wavefront methods \cite{Wolfe1986,2000:UTS:846234.849346}, and their combinations with various tile shapes such as diamond and hexagonal tiling \cite{wavefrontdiamond,doi:10.1142/S0129626414410023}. 

The only works we are aware of that has applied similar transformations to large-scale scientific problems are the Formura DSL \cite{Muranushi}, which is in full control of the code that is being generated from high-level mathematical expressions - therefore it avoids the issue of various execution paths and multiple compilation units to tile across. Work by Malas et. al \cite{7516010} applied a combination of wavefront and diamond tiling to an electromagnetics code, however only a handful of loops are tiled across, and it is done mostly by hand.

A common point in all of the above research is that the transformations are applied at compile-time (or before), and therefore they are inherently limited by what is known at compile time, and the scope of the analysis. This in turn makes their application to large-scale codes distributed across many compilation units, that have configurable, complex execution flows and call stacks, exceedingly difficult.

Identifying the sequence of loops to tile across and to carry out dependency analysis is a lot easier at run-time, particularly with the help of delayed evaluation or lazy execution \cite{Henderson:1976:LE:800168.811543,bloss1988code}, which is a well-known technique used particularly in functional languages that allows expressions to be evaluated only when their results are required. Lazy execution is also used in other fields, such as Apache Spark to plan out the sequence of computations and to skip unnecessary steps. We apply the lazy execution idea to figure out dependencies and compute loops schedules at runtime - to our knowledge these two have not been used together in scientific computing.

\begin{figure}\centering
\includegraphics[width=0.40\textwidth]{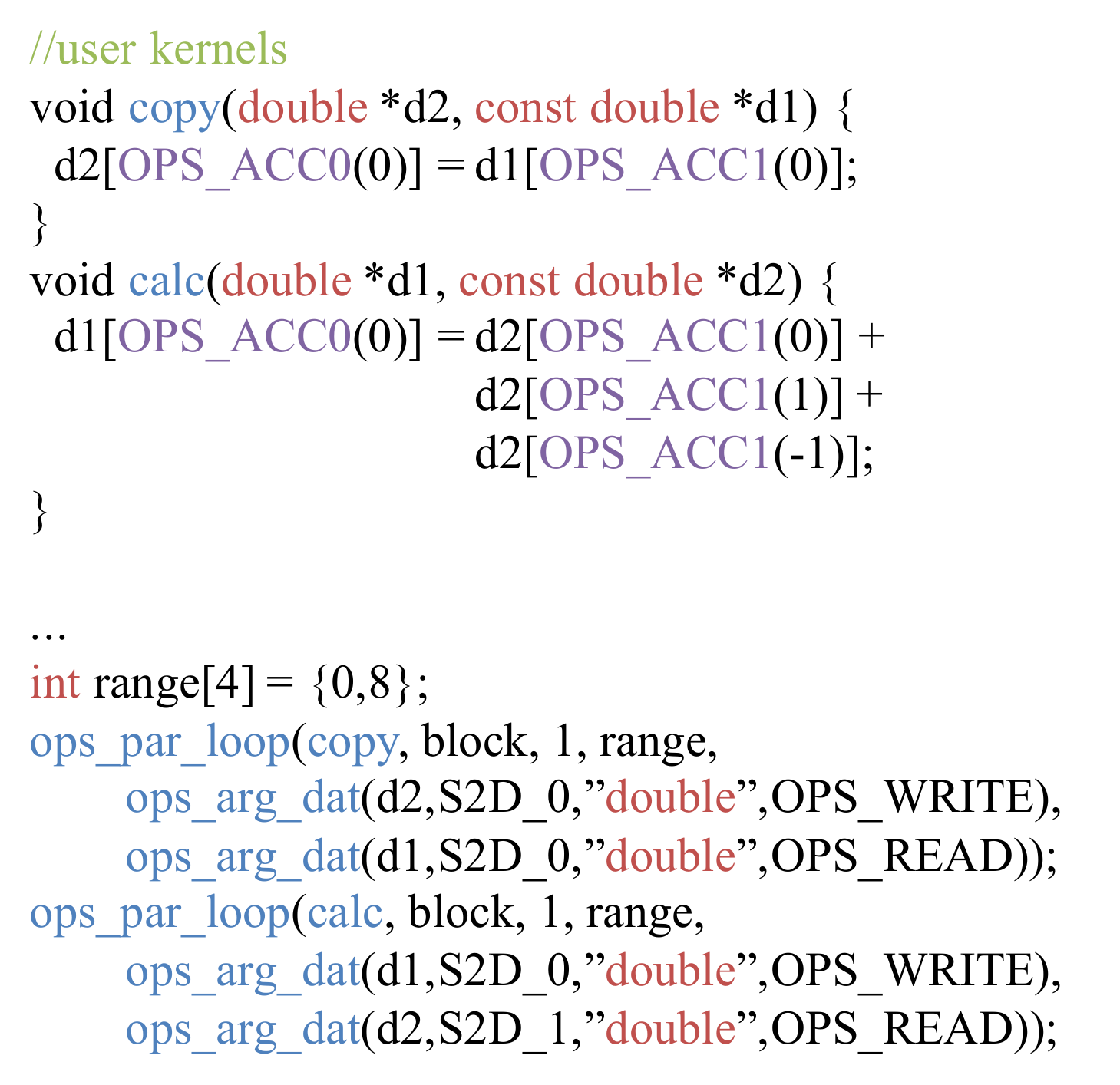}\vspace{-10pt}
\caption{\small An OPS parallel loop\normalsize}\vspace{-15pt}\label{fig/parloop}
\end{figure}

\vspace{-10pt}
\section{The OPS Embedded DSL}
\label{sec/ops}

The Oxford Parallel library for Structured meshes (OPS) is a Domain Specific Language embedded into C/C++/Fortran, defining an API for expressing computations on multi-block structured meshes. 

It can be used to express algorithms at a higher level, without having to worry about the intricacies of parallel programming and data movement on various computer architectures. By separating the high-level code from the low-level implementation, OPS lets domain scientists to write a single high-level source code and the library developers to automate the generation of low-level implementations given the knowledge of the domain and the target architectures.

OPS defines the following abstraction: the computational domain consists of a number of N dimensional \emph{blocks}, with a number of \emph{datasets} defined on each. Then, following the access-execute model \cite{aecute}, computations are expressed as a sequence of parallel loops applying given ``user-kernels'' over given iteration ranges and a number of datasets defined on the same block, specifying how each dataset is accessed: whether it is read, written, or incremented and what exact stencil is used for the access. This requires the parallel operation to be insensitive to the order of execution on individual grid points (within machine precision).

An example of an OPS parallel loop is shown in Figure \ref{fig/parloop}; the \texttt{ops\_par\_loop} API call takes as arguments a function pointer to be applied to each grid point, a block, a dimensionality, an iteration range and a number of data arguments. A data argument encapsulates the dataset handle, the stencil, the underlying primitive datatype and the type of access. 

Given this abstraction, OPS is free to parallelise both over parallel loops over different blocks, as well as over individual grid points within a single parallel loop: indeed the library assumes responsibility for correctly parallelising in distributed-memory as well as shared-memory environments, and on different architectures, using different parallel programming models. 
With a user code written once using the C/C++ or Fortran API of OPS, a source-to-source translator generates code for sequential, OpenMP, OpenACC, OpenCL and CUDA execution, which is then compiled with a traditional compiler and linked against one of the OPS back-end libraries that supports MPI parallelisation and data management. Because ownership of data is handed to the library, and access only happens through OPS APIs, the library can keep track of what data changed and when it is necessary to update it: halos for MPI or the separate address spaces of CPUs and GPUs. 

\section{Skewed tiling algorithm}
\label{sec/alg}

As described in the previous section, at runtime the \texttt{ops\_par\_loop} construct includes all necessary information about a computational loop that is required to execute it: the computational kernel, the iteration range, and a list of datasets, plus how they are accessed - the stencil and whether read or written. This enables OPS to store this information for delayed execution, and reason about multiple loops - following the loop chaining abstraction \cite{6877342}.

\subsection{Delayed execution}

With all pertinent information about a loop, we create a C struct at runtime, which includes a function pointer to a C++ function that, given the loop ranges and the argument list stored in the struct, can execute the computational loop. When the \texttt{ops\_par\_loop} is called from user code, this struct is passed to the back-end, and stored in an array for later execution. Parallel loops can be queued up until the point when the user code needs some data to be returned: such as getting the result of a reduction, based on which a control decision has to be made. At this point, OPS triggers the execution of all loops in the queue.

\subsection{Dependency analysis}

Having queued up a number of computational loops, it is now possible to carry out dependency analysis: this enables us to reason about loop scheduling not only in individual loops but across a number of loops as well. The ultimate goal is to come up with execution schedules that dramatically improve data locality by way of cross-loop blocking (also called sparse tiling). Therefore, the dependency analysis carried out by OPS takes into consideration the sequence of loops, the datasets accessed by each loop, the stencils used and whether the data is read, written, or both. Given the restrictions of the OPS abstraction (only trivially parallel loops permitted), the run-time information about datasets and stencils used to access them, the dependency analysis is based on the well-known polyhedral model.


As the theory of transformations to polyhedral models is well documented \cite{xue2012loop,Bondhugula}, here we focus only on the overall algorithmic description and some practicalities. Unlike many algorithms, we do not assume nor exploit any recurrence, the time dimension is simply replaced by a sequence of loops that may all have different iteration ranges and stencils. In this paper, we use dependency analysis to implement a skewed tiling scheme, with a sequential dependency (and scheduling) between subsequent tiles and intra-tile parallelism.

\begin{figure}[t!]\centering
\includegraphics[width=0.5\textwidth]{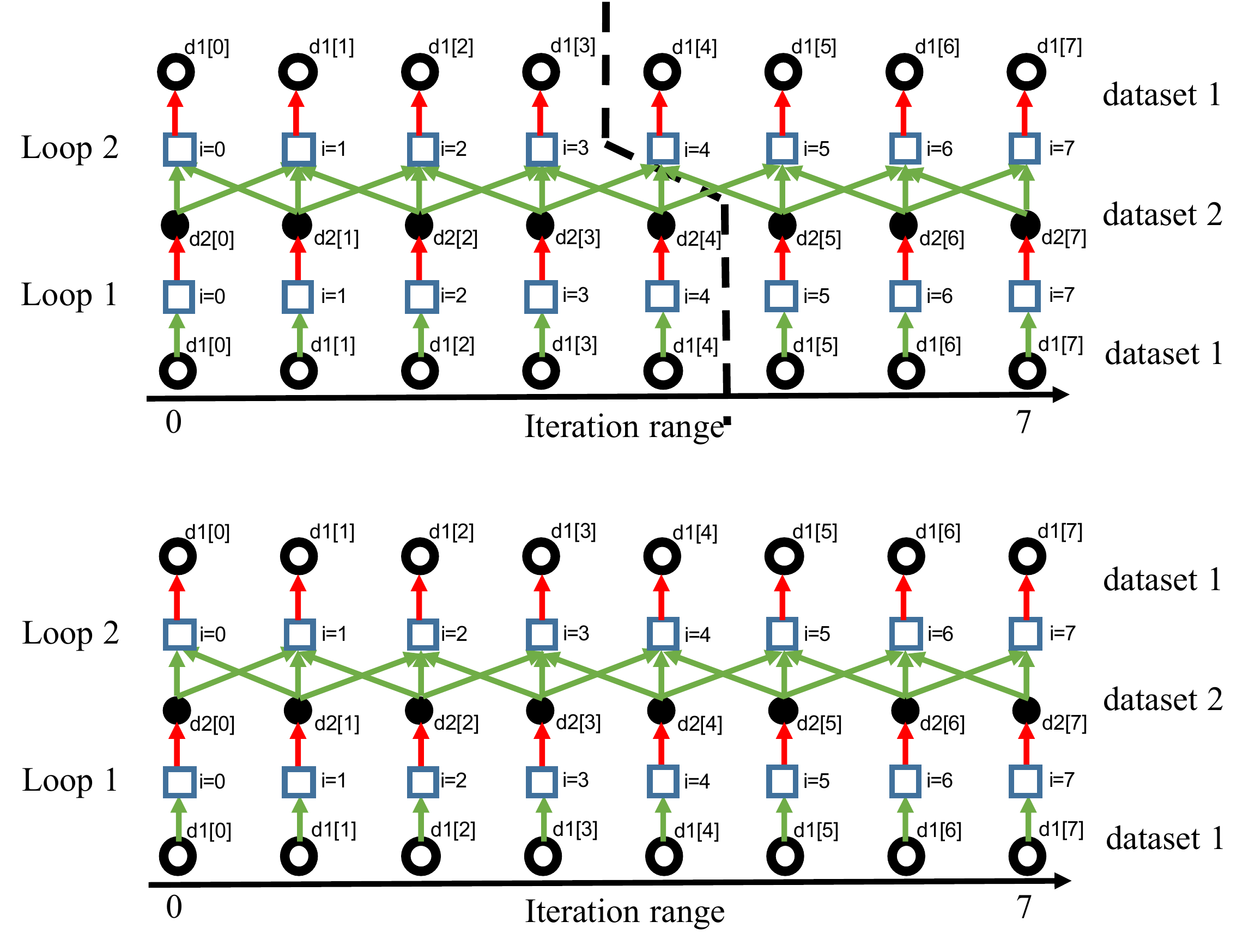}\vspace{-10pt}
\caption{\small Illustrative example of 1D dependency analysis and tiling\normalsize}\vspace{-15pt}\label{fig/tiling1d}
\end{figure}

The key idea to understand tiling can be demonstrated on a simple 1D example with two loops that are shown in Figure \ref{fig/parloop}; the first reading dataset 1 and writing dataset 2, the second reading dataset 2 with a 3-point stencil, and writing dataset 1, as illustrated in Figure \ref{fig/tiling1d}. This is the conventional way of formulating computational loops; the issue is that these datasets are too large to fit in the on-chip cache, they will be streamed to and from the CPU between loops. This implies that data re-use only happens within a single loop nest, and not across subsequent loop nests. Data re-use in this case is 3x of the values of dataset 2 in the second loop (to have computations and data movement in balance on modern CPUs, one would need a factor of $\sim40\times$).

In order to improve locality and re-use data across different loop nests, we need to take multiple loop nests and reason about them together; we would like to apply loop blocking within and across the two loops (tile across the two loops). To achieve this, we need to block the iteration ranges of loops and re-organise them so that data accessed by a given block in the first loop stays in cache and the second loop then executes on that same block. Then, we execute the next block on the first loop, then on the second loop, keeping data in cache in-between. Thus, we can achieve cross-loop data re-use. The caveat is that in constructing these blocks, we need to make sure all data dependencies are satisfied; only those iterations can be included in the block for the second loop, for which the block of the first loop computed the required inputs.

One way to partition the index spaces of loops 1 and 2, is to first partition the iteration range of loop 2 into two equal parts; $[0-3]$ for Tile 1 and $[4-7]$ for Tile 2. Then we analyse loop 2 to determine its data dependencies: in order to be able to execute iterations $[0-3]$ of loop 2, we need up-to-date values of dataset 2 on grid points $[0-4]$. Stepping backward to loop 1, we need to make sure that we set its iteration range in a way that satisfies all data dependencies: as loop 1 is responsible for writing dataset 2, the iteration range for Tile 1 becomes $[0-4]$, wider than that of loop 2: this is called skewing. The iteration range for Tile 2 on loop 1 is what is left over, $[5-7]$, despite its data dependency over indices $[3-7]$ of dataset 2: we rely on Tile 1 computing indices $[3-4]$ before the execution of Tile 2 is started, introducing a serial dependency between tiles. Now, relevant parts of datasets 1 and 2 can stay in cache across the execution of loop 1 and 2, further improving data re-use.

This method can be generalised to work in arbitrary dimensions and with a much larger number of loops as below. While it is possible to determine dependencies and the required skew across loops based on just the stencil patterns, we carry out dependency analysis within each tile across individual loops as well, as it makes handling corner cases easier. We first give an informal description of the 7-step process creating tiling plans or execution schedules, and then give the specific algorithm use in OPS.

\begin{enumerate}
	\item Lines 1-6: First, we determine the union of all iteration ranges, and partition it into N tiles - the exact size is a configurable parameter.
	\item Line 7-8: Looping over the sequence of computational loops in reverse order, we loop over each dimension and each tile:
	\item Lines 9-14: Start index for the current loop, in the current dimension, for the current tile, is either the end index of the previous tile, or the start index of the original index set.
	\item Lines 15-24: End index is calculated based on a read dependency of a loop with a higher index in the tile for any datasets written.
	\item Lines 28-33: End index updated to account for write-after-read and write-after-write dependencies across tiles where the ordering will effectively change\footnote{This is to avoid reading a value that was written by a higher index loop in a previous tile - further skewing has to be introduced to the lower index loop in the tile that writes the value in a higher index loop.}.
	\item Lines 35-39: If no read or write dependencies exist, the end index is the minimum of the end index of the original iteration range for this loop and the start index plus the tile size from step 1.
	\item Lines 40-53: Finally, based on the computed iteration range, the read and write dependencies of datasets are updated, accounting for the stencils used.
\end{enumerate}

More formally, the algorithm is given a set of loops $loop_l$, with iteration ranges in each dimension $loop_l.start_d$, $loop_l.end_d$, a number of arguments and stencils. Our goal is to construct a number of tiles, each containing iteration ranges for each loop $tile_{t_d}.loop_l.start_d$, $tile_{t_d}.loop_l.end_d$, in a way that allows for the re-organisation of execution in a tiled way:

\begin{algorithmic}[1]
\STATE \COMMENT{ compute union of index sets}
\FORALL{l in loops, d in dimensions}
\STATE $start_d = min(start_d,loop_l.start_d)$
\STATE $end_d = max(end_d,loop_l.end_d)$
\STATE $num\_tiles_d = (end_d-start_d-1)/tilesize_d+1$
\ENDFOR
\FORALL{l in loops backward, d in dimensions}
\FORALL{t in tiles}
\STATE \COMMENT{ start index for current loop, dimension and tile}
\IF{t first in d}
\STATE $tile_{t_d}.loop_l.start_d = loop_l.start_d$
\ELSE
\STATE $tile_{t_d}.loop_l.start_d = tile_{{t_d}-1}.loop_l.start_d$
\ENDIF
\STATE \COMMENT{ end index for current loop, dimension and tile}
\STATE $tile_{t_d}.loop_l.end_d = -\infty$
\IF{t last non-empty in d}
\STATE $tile_{t_d}.loop_l.end_d = loop_l.end_d$
\ELSE
\STATE \COMMENT{ satisfy read-after-write dependencies }
\FORALL{a in arguments of loop written}
\STATE $tile_{t_d}.loop_l.end_d = min(loop_l.end_d, $
\STATE $\quad max(tile_{t_d}.loop_l.end_d,read\_dep_a.tile_t.end_d))$
\ENDFOR
\ENDIF
\ENDFOR \COMMENT{ over tiles }
\FORALL{t in tiles}
\IF{t not last}
\STATE \COMMENT{ satisfy write-after-read/write dependencies }
\FORALL{a in arguments of loop}
\STATE $m =$ largest negative stencil point in $d$
\STATE $tile_{t_d}.loop_l.end_d = min(loop_l.end_d, max($\\$tile_{t_d}.loop_l.end_d,write\_dep_a.tile_t.end_d-m))$
\ENDFOR
\ENDIF
\STATE \COMMENT{ default to end index at tile size }
\IF{$tile_{t_d}.loop_l.end_d$ still $-\infty $}
\STATE $tile_{t_d}.loop_l.end_d = min(loop_l.end_d,$
\STATE $\qquad start_d+t_d*tilesize_d)$
\ENDIF
\STATE \COMMENT{ update read dependencies }
\FORALL{a in arguments of loop read}
\STATE $p =$ largest positive stencil point in $d$
\STATE $read\_dep_a.tile_t.end_d = $\\$max(read\_dep_a.tile_t.end_d, tile_{t_d}.loop_l.end_d+p)$
\STATE $p =$ largest negative stencil point in $d$
\STATE $read\_dep_a.tile_t.start_d = $\\$min(read\_dep_a.tile_t.start_d, tile_{t_d}.loop_l.start_d+p)$
\ENDFOR
\STATE \COMMENT{ update write dependencies }
\FORALL{a in arguments of loop written}
\STATE $write\_dep_a.tile_l.end_d = $
\STATE $\quad max(write\_dep_a.tile_t.end_d,tile_{t_d}.loop_l.end_d)$
\STATE $write\_dep_a.tile_l.start_d = $
\STATE $\quad min(write\_dep_a.tile_t.start_d,tile_{t_d}.loop_l.start_d)$
\ENDFOR
\ENDFOR
\ENDFOR
\end{algorithmic}

The algorithm produces as its output the iteration ranges for each loop in each tile - this is then cached as a ``tiling plan'' and re-used when the same sequence of loops is encountered. Given a tiling plan, the execution of the tiled loop and iteration schedule is described by the following algorithm: we iterate through every tile, and subsequent loops, replacing the original iteration range with the range specific to the current tile (loops with empty index sets are skipped), then start execution through the function pointer. Parallelisation happens within the tiles.

\begin{algorithmic}[1]
\FORALL{tiles t=1..T}
\FORALL{loops l=1..L}
\FORALL{dimensions d=1..D}
\STATE $bounds\_start_d = tile_{t_d}.loop_l.start_d$
\STATE $bounds\_end_d = tile_{t_d}.loop_l.start_d$ 
\ENDFOR
\STATE call $loop_l$ with bounds: 
\STATE \qquad $bounds\_start_d,bounds\_end_d$
\ENDFOR
\ENDFOR
\end{algorithmic}

For this work, we chose the skewed tiling algorithm instead of more advanced algorithms such as diamond tiling \cite{diamond} or its combination with wavefront scheduling \cite{wavefrontdiamond}, because it is both simpler to implement and verify, and it results in large tiles, helping to diminish the overhead of launching the execution of computational loops through function pointers (detailed in section \ref{sec/heat}). OPS however captures all the information necessary to apply more complex loop scheduling, which will be the target of future research.

\subsection{Tile size selection}

We designed an algorithm in OPS to automatically choose a tile size based on the number of datasets and the size of the last-level cache: by looking at the chain of loops to be tiled across, it counts the number of datasets accessed, adds their size up to compute the total amount of data per grid point, then produces a tile shape that will fit in cache and the X size is at least twice the Y size, and the product of Y size and Z size is an integer multiple of the number of threads (considering that we parallelise within the tiles). To demonstrate performance at different tile sizes, and the best achieved performance, later in the results section we will show results at the best tile size found manually, and we also indicate the performance achieved by the automatic tile selection algorithm. An exploration of possible tile sizes is deferred to the Supplementary Material.

\subsection{Extension to distributed memory systems}

The parallel loop abstraction of OPS lets it deploy different kinds of parallelisation approaches, including support for distributed memory systems through the Message Passing Interface (MPI). Without any additional user code, OPS will automatically perform a domain decomposition, create halo regions for all datasets, and given the stencil access patterns in \texttt{ops\_par\_loop} constructs, automatically keep the values of the halo up-to-date. The performance and scalability of OPS has been presented and analysed in previous work \cite{ops2}.

In deploying cache blocking tiling to large scale stencil codes, the obvious next step is to introduce support for tiling in distributed memory systems. Given that OPS is in full control of data and parallel execution, as well as the MPI decomposition and scheduling of computations and communications, it is again possible to do this without any user intervention.

In a distributed memory system, the two key issues in tiling are the construction and scheduling of tiles, and the communication of data required for executing the tiles. Our skewed tiling approach in shared memory systems introduces a sequential dependency across tiles, which prohibits parallelisation between tiles (there we only parallelise within tiles) - this is obviously not a viable approach over MPI. Instead, we apply an overlapped tiling approach \cite{overlap}, where points in the iteration space along the boundaries of the domain decomposition that are required for the execution of tiles on the other side of the boundary are replicated there. Thus, these iterations will be computed redundantly on an MPI process that does not own them, requiring communicating the needed data on those points.

In terms of the construction of tiles this means that some will extend beyond the original boundaries of domain decomposition, but in terms of scheduling it also means that MPI processes can execute their tiles independently (and in parallel) of one another. In terms of communications, the halo regions are extended to accommodate data required for the computation of these iterations, and before executing the tiles, MPI processes exchange a wider halo of datasets that are read before they are written. During the execution of the tiles however there is no need for further communication, because for each tile all data required is in local memory. This is in contrast to the existing MPI communications scheme in OPS, which is on-demand: halos are updated immediately before loops where they will be accessed, and there are no redundant computations.

\begin{figure}[t!]\centering
\includegraphics[width=0.45\textwidth]{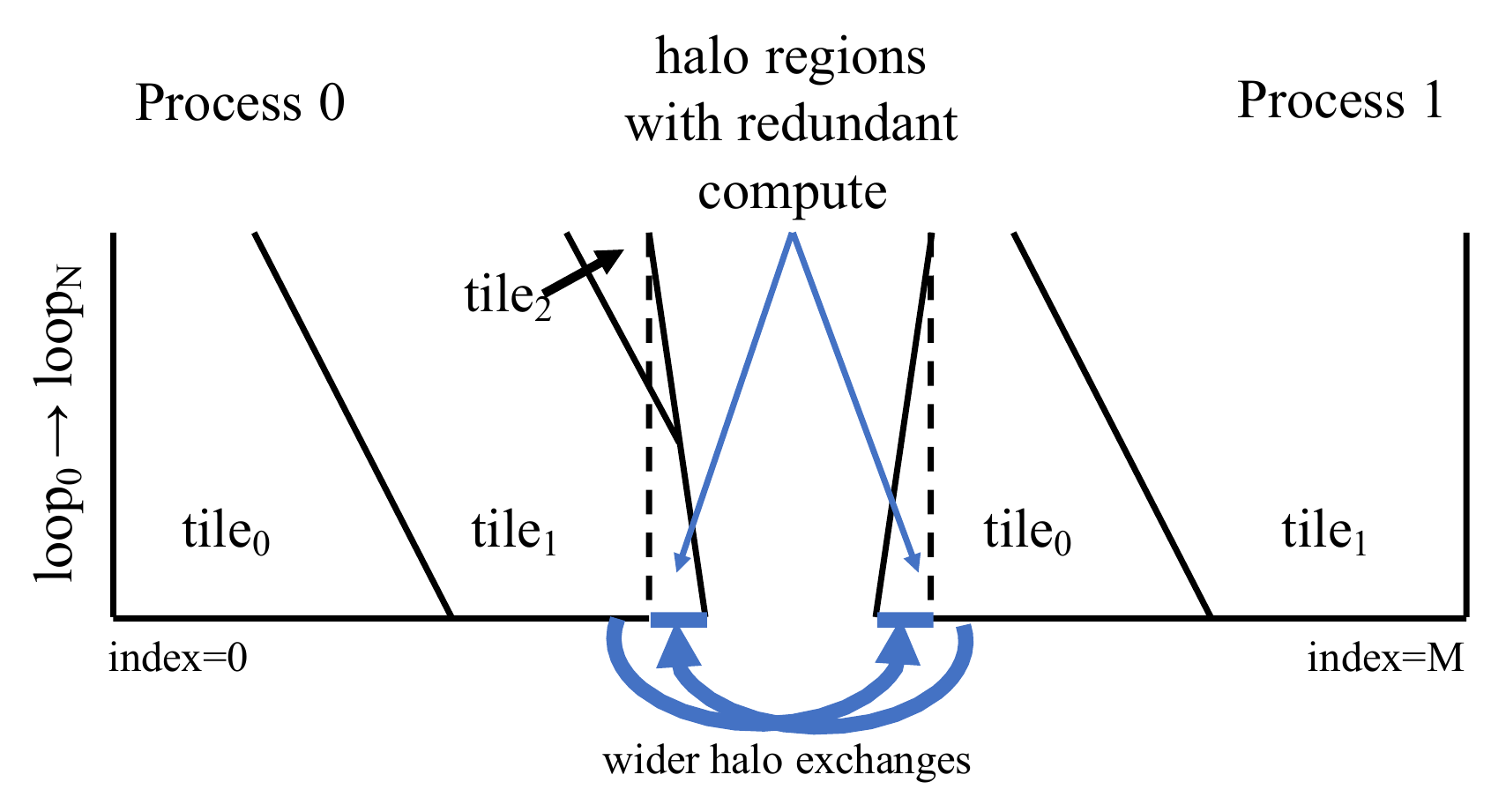}\vspace{-15pt}
\caption{\small A 1D illustrative example of tiling across MPI partitions\normalsize}\vspace{-15pt}\label{fig/tiling_mpi}
\end{figure}

There are two key changes to the construction of tiles and the execution schedule, as illustrated on Figure \ref{fig/tiling_mpi}. First, the calculation of $tile_{t_d}.loop_l.start_d$ and $tile_{t_d}.loop_l.end_d$ has to account for domain decomposition boundaries: to compute the first tile's start and the last tile's end index, we need to extend beyond the boundaries, but we only have to look at read dependencies (no need to look at write dependencies in overlapped tiling). Second, we have to compute the required halo depth to communicate.

Before either of these, a minor modification to compute the number of tiles on each process: lines 3-4 of the original algorithm change to calculate with the bounds given by the domain decomposition:
\begin{algorithmic}
\STATE \COMMENT{line 3-4}
\STATE $start_d = min(start_d,loop_l.start\_thisprocess_d)$
\STATE $end_d = min(end_d,loop_l.end\_thisprocess_d)$
\STATE
\end{algorithmic}

\vspace{-15pt}
\subsubsection{Tile shapes at MPI boundaries}

When calculating the starting index for the first tile on the process, we have two separate cases: (1) when the process has no ``left'' neighbour in the current dimension (Figure \ref{fig/tiling_mpi}, $tile_0$ of Process 0) , then the tile's start index is just the start index of the original loop bounds as previously (line 11). (2) when the process has a ``left'' neighbour: then the iteration range has to be further extended to the left to account for read dependencies (Figure \ref{fig/tiling_mpi}, $tile_0$ of Process 1). This is described by the following algorithm, which replaces line 11 of the original:
\begin{algorithmic}
\STATE \COMMENT{line 11}
\IF{process is first in d}
\STATE $tile_{t_d}.loop_l.start_d = loop_l.start_d$
\ELSE
\STATE \COMMENT{ satisfy read-after-write dependencies }
\FORALL{a in arguments of loop written}
\STATE $tile_{t_d}.loop_l.start_d = max(loop_l.start_d, $
\STATE $\qquad min(tile_{t_d}.loop_l.start_d,read\_dep_a.tile_t.start_d))$
\ENDFOR
\ENDIF
\end{algorithmic}

When calculating the end index of the last tile, we only account for read-after-write dependencies, which now will extend beyond the MPI partition boundary, due to the logic in line 22 (Figure \ref{fig/tiling_mpi}, $tile_1$ and $tile_2$ of Process 0). Consider, that the slope of the two sides of the last tile is asymmetric: the left edge (which is the right edge of the previous tile) has to consider both read and write dependencies, but the right edge only considers read dependencies. In particularly long but slim tiles this may lead to the left edge reaching the right edge, and the tile having no iteration space in early loops, as illustrated by $tile_2$ on Figure \ref{fig/tiling_mpi}. In accounting for this possibility, we have to check for this ``overshoot'' and if it does happen, then the previous tile's end index will have to be adjusted. In our algorithm, we have to include two checks, thus lines 28-34 are changed as follows:

\begin{algorithmic}
\STATE \COMMENT{line 28-34}
\IF{t not last and $tile_{t_d}.loop_l.end_d > tile_{t_d}.loop_l.start_d$}
\STATE \COMMENT{ satisfy write-after-read/write dependencies }
\FORALL{a in arguments of loop}
\STATE $m =$ largest negative stencil point in $d$
\STATE $tile_{t_d}.loop_l.end_d = min(loop_l.end_d, max($\\$tile_{t_d}.loop_l.end_d,write\_dep_a.tile_t.end_d-m))$
\ENDFOR
\STATE \COMMENT{ if we overshoot the next tile }
\IF{$tile_{t_d}.loop_l.end_d > tile_{t_d+1}.loop_l.end_d$}
\STATE $tile_{t_d}.loop_l.end_d = tile_{t_d+1}.loop_l.end_d$
\ENDIF
\ENDIF
\end{algorithmic}
\vspace{-3pt}
\subsubsection{Halo exchanges}

Determining which datasets require a halo exchange and what depth is very simple given the correctly populated $read\_dep_a.tile_t.start_d, read\_dep_a.tile_t.end_d$ arrays; for each dataset we look for the first loop that accesses it, if it is a write, then no halo exchange is necessary, if it is a read, then the depth to exchange is the difference between the domain decomposition boundary and the read dependency index of the first/last tile, as illustrated on Figure \ref{fig/tiling_mpi}. Since exchange depth may not be symmetrical, each process keeps track of the read dependencies of its previous neighbour's last tile and its next neighbour's first tile.

This information is saved alongside the tiling plan, and each time we encounter the same sequence of loops, we first carry out these halo exchanges, then execute the tiling plan itself, which then does not need any further MPI communications until its completion.

\section{Stencil Applications}
\label{sec/apps}

In order to evaluate the efficiency of cache-blocking tiling in OPS, we first carry out an in-depth study using one of the most commonly used benchmarks in tiling research: a Jacobi iteration solving the heat equation. \emph{We should note that our tiling results on the Jacobi benchmark are in no way new, they only serve to establish a baseline, and to help understand the performance implications of our run-time tiling approach.}

There are two computational stages in the Jacobi benchmark: apply a 5-point weighted finite difference stencil, and then copy the results back to the original array.

This benchmark problem is implemented in both Pluto and Pochoir, and comes as part of the distributed source package, and we use it as a basis for comparison. Pochoir's implementation differs from Pluto's in that it unrolls the time iteration thereby avoiding the second step, copying $b$ to $a$ (denoted as copy version); rather it applies the stencil to $a$ first, putting the result in $b$, then the other way around, which is then repeated for the desired number of iterations (denoted as non-copy version).

Our main benchmark application is CloverLeaf \cite{cloverleaf}; to our knowledge, successful cache-blocking tiling has not been previously reported on an application this size. CloverLeaf is a mini-application that solves the compressible Euler equations on a Cartesian grid, using an explicit second-order method.
It uses a Lagrangian-Eulerian scheme to solve Euler's equations for the conservation of mass, energy and momentum, supplemented by an ideal gas equation. CloverLeaf uses an explicit time-marching scheme, computing energy, density, pressure and velocity on a staggered grid, using the finite volume discretisation. One timestep involves two main computational stages: a Lagrangian step with a predictor-corrector scheme, advancing time, and an advection step - with separate sweeps in the horizontal/vertical/depth dimensions. The full source of the original is available at \cite{cloverleaf-github}. 

CloverLeaf was an ideal candidate for porting to use the OPS library - indeed it is the first application that was developed for OPS \cite{ops2}. The 2D/3D application consists of 25/30 datasets defined on the full computational domain (200/240 bytes per grid point), and 30/46 different stencils used to access them. There are a total of 83/141 parallel loops spread across 15 source files, each using different datasets, stencils and ``user kernels''; many of these include branching (such as upwind/downwind schemes, dependent on data). Furthermore, the source files that contain \texttt{ops\_par\_loop} calls include branching themselves, dependent on e.g. sweep direction, with some code paths shared and some different for different sweeps, and often the pointers used refer to different datasets, depending on the call stack. A single time iteration consists of a chain of 153/603 subsequent loops. The full size of the CloverLeaf is 4800/6000 lines of code.

These properties of CloverLeaf make it virtually impossible to apply stencil compilers as they are limited by what is known at compile-time - which is indeed very little for larger-scale codes. While some portions of the code (blocks of 4-5 consecutive loops) are amenable to compile-time tiling approaches, there is little data reuse for tiling to show any performance benefit (as experiments with Pluto showed). This motivates our research into tiling with OPS that is capable of constructing and executing tiles at run-time.

To show that our results generalise to other applications using OPS, we also briefly evaluate performance on two more applications. The first is TeaLeaf 2D, also part of the Mantevo uite, which is a matrix-free sparse linear solver code for hydrodynamics applications. TeaLeaf has 98 nested loops over the 2D grid, 31 datasets defined on the grid, and the code is spread across 12 source files. It has a wide range of configuration parameters that control its execution: it supports various algorithms including Conjugate Gradient (CG), Chebyshev, and Preconditioned Polynomial CG (PPCG). At runtime, depending on the level of convergence and various problem-specific parameters, it will perform different numbers of preconditioning iterations, has early exits and other control structures that make it particularly unsuitable for polyhedral compilers.

The fourth key OPS application is OpenSBLI \cite{opensbli}, a large-scale academic research code being developed at the University of Southampton, focusing on the solution of the compressible Navier-Stokes equations with application to shock-boundary layer interactions (SBLI). Here we are evaluating a 3D Taylor-Green vortex testcase, which consists of 27 nested loops over the computational grid, using 9 different stencils and accessing 29 datasets defined on the 3D grid.

\vspace{-5pt}
\section{Benchmarking and performance analysis}
\label{sec/bench}
\vspace{-3pt}
\subsection{Experimental set-up}
\vspace{-2pt}
We evaluate all algorithms and codes on a dual-socket Intel Xeon E5-2650 v3 (Haswell) machine, that has 10 physical cores and 20 MB of L3 cache per socket. Hyper-Threading is enabled. For all tests, we run on a single CPU socket using \texttt{numactl} in order to avoid any NUMA effects, and parallelise with OpenMP within tiles (20 threads, pinned to cores). The latest version of OPS is available at \cite{ops-github}. We use Pluto 0.11.4 (dated Oct 28, 2015) and a Pochoir version dated Apr 15, 2015, available from GitHub. All codes are compiled with the Intel compilers version 17.0.3, with \texttt{-fp-model fast} and fused multiply-adds enabled.

For a simple roofline model, we benchmark a single socket of the system. Achieved bandwidth to DDR4 memory is 49 GB/s using the STREAM benchmark (Triad, 50M array, repeated 100 times), and 227 GB/s bandwidth to L3 cache (Triad, 900K array, repeated 1000 times). The double precision general matrix-matrix multiply test, using MKL, shows an achieved peak computational throughput of 270 GFLOPS/s. We use these figures for later analysis: with the balance point between computations and communication at 44 Flop/DWord - below the performance is bound by bandwidth, above it is bound by compute throughput.

Bandwidth figures shown in the following are based on back of the envelope calculations, ignoring data re-use within a single loop due to multi-point stencils; these values stay in cache. In the case of CloverLeaf we use the automated reporting system in OPS that estimates bandwidth based on the iteration range and the type of access (R/W) to data - for this calculation the data re-use due to multi-point stencils is ignored. As the Haswell microarchitecture does not have counters for floating point operations, we use counters in NVIDIA GPUs - we run OPS CUDA variants of our applications (the computational kernels are identical to the CPU implementation) through the \texttt{nvprof} profiler collecting double-precision flop counters using \texttt{--metrics flop\_count\_dp}. These flops counts are then used as they are to estimate  GFLOPS/s throughput figures on the CPU.

\vspace{-5pt}
\subsection{Heat equation} \label{sec/heat}
As one of the most studied examples for tiling, we carry out the analysis on a Jacobi iteration solving the 2D heat equation. We solve on a $8192^2$ mesh, with one extra layer for a Dirichlet boundary condition on all sides, for 250 time iterations. All data and computations are in double precision, and the total memory footprint is 1 GB.


\vspace{-5pt}
\subsubsection{Pluto}
First, we evaluate Pluto, compiling with the recommended flags: \texttt{./polycc test/jacobi-2d-imper.c --tile --parallel}, and running a series of tests at different tile sizes. Without tiling (with OpenMP), the runtime is 16.07 seconds, achieving 31.1 GB/s. Pluto constructs a number of tiles in sets that are inter-dependent, and parallelises over different tiles in the same set. The best performance is achieved at an X tile size of 192, a Y tile size of 32, tiled over 15 time iterations, which corresponds to a base tile footprint of 12 KB per thread. The test completes in 3.5 seconds, achieving 142 GB/s. In terms of computational throughput, this corresponds to 33.5 GFLOPS/s.
\vspace{-5pt}
\subsubsection{Pochoir}

Second, we evaluate Pochoir - as previously described, its implementation avoids straight copies from one array to another, therefore it is slightly faster. Without tiling, the reference implementation runs in 9.4 seconds, achieving 26.6 GB/s. Pochoir, similar to Pluto, parallelises over different tiles, the \texttt{heat\_2D\_NP} version runs in 3.26 seconds and achieves 76.7 GB/s and 36 GFLOPS/s, and the zero-padded version \texttt{heat\_2D\_NP\_zero} runs in 2.92 seconds, and achieves 85.6 GB/s and 40.5 GFLOPS/s. 
\vspace{-5pt}
\subsubsection{Hand-coded benchmarks}
Third, we implement small experimental codes that solve the heat equation, one making the copies, the other avoiding them. In a similar way to Pluto, the tile sizes are known at compile time, however, (unlike Pluto) the mesh size and the number of iterations are not. The key difference between this code and Pluto/Pochoir, is that it uses the same sort of skewed tiles that OPS does, and parallelises within the tile. For this benchmark, we only tile in time and the Y dimension, and not in the X dimension (which is equivalent to choosing an X tile size of 8192). The baseline performance without tiling is 8.31 seconds, or 30.1 GB/s.
With tiling, the best performance is achieved at a Y tile size of 120, tiling over 50 time iterations 2.43 seconds or 101 GB/s and 48 GFLOPS/s - memory used is 15MB (vs. 20MB of L3) per tile. The copy version of the code (that is similar to Pluto), brings the baseline performance to 16.36 seconds or 30.4 GB/s, and the best tiled performance is achieved at a Y tile size of 160, tiling over 50 time iterations, with a runtime of 3.54 seconds, achieving 141.2 GB/s or 33.1 GFLOPS/s.

Next, we outline the computational loop in our hand-coded benchmark into a separate source file (accepting data pointers and the X,Y iteration ranges as arguments), so as to simulate the way OPS calls the computational code. We compile with inter-procedural optimisations turned off to make sure the function does not get inlined - this reduces performance by 30-45\% across the board, bringing the best performance of the non-copy variant from 2.43 to 3.54 seconds, or 70.1 GB/s. The performance of the hand-coded copy variant goes from 3.54 to 4.7 seconds, or 105.6 GB/s. This establishes an upper bound for performance that is achievable through OPS, where computational subroutines are outlined and are in separate compilation units, no compile-time tile size or alignment information is available.
\vspace{-5pt}
\subsubsection{OPS}
Finally, we evaluate performance in OPS. Note, that OPS has the least amount knowledge of compile-time parameters, and it uses a completely generic dependency analysis at run-time, as opposed to all previous tests which do the dependency analysis at compile-time. The Jacobi iteration is implemented both ways, copy and non-copy. Without tiling, the runtime of the copy variant is 16.06 seconds, achieving 31.1 GB/s. The non-copy variant runs in 8.58 seconds, achieving 29.4 GB/s. 
After switching on tiling and tuning the tile size, best performance is achieved at 8192 X size, 100 Y tile size, tiled over 30 time iterations, with a runtime of 5.11 seconds, achieving 98.8 GB/s or 22.9 GFLOPS/s - 10\% slower than the outlined hand-coded benchmark. The non-copy variant runs in 3.69 seconds, achieving 67.3 GB/s, 5\% slower than the outlined hand-coded benchmark. The overhead of computing the tiled execution scheme was 0.0042 seconds for the copy, and 0.0040 seconds for the non-copy version - 0.1\% or less of the total runtime.

It is important to observe that in all of the above benchmarks, the optimal performance was achieved when the X tile size was considerably bigger than the Y tile size - this is due to the higher efficiency of vectorised execution and prefetching: X loops are peeled by the compiler to get up to alignment with non-vectorised iterations, then the bulk of iterations are being vectorised over, and finally there are scalar remainder iterations.

\begin{table}\small
\centering
\caption{Performance summary of the copy Jacobi iterations\normalsize}
\vspace{-6pt}\small
\begin{tabular}{ l | c | c | r }
  \hline			
  Test & Type & Baseline (s) & Tiled (s) \\ \hline \hline
  Pluto   & copy & 16.07 & 3.50 \\ \hline
  hand-coded & copy & 16.36 & 3.54 \\ \hline
  OPS & copy & 16.06 & 5.11 \\ \hline
\end{tabular}\label{tab/jacobi_summary}
\vspace{-10pt}
\end{table}\normalsize

\begin{table}\small
\centering
\caption{Performance summary of the non-copy Jacobi iterations\normalsize}
\vspace{-6pt}\small
\begin{tabular}{ l | c | c | r }
  \hline			
  Test & Type & Baseline (s) & Tiled (s) \\ \hline \hline
  Pochoir & non-copy & 9.4 & 2.92 \\ \hline
  hand-coded & non-copy & 8.31 & 2.43 \\ \hline
  OPS & non-copy & 8.58 & 3.69 \\ \hline
\end{tabular}\label{tab/jacobi_summary2}
\vspace{-10pt}
\end{table}\normalsize

\subsubsection{Comparison of tiling implementations}
Overall, as summarised in Tables \ref{tab/jacobi_summary} and \ref{tab/jacobi_summary2}, the performance of non-tiled versions is quite consistent: copy variants (Pluto, hand-coded, and OPS) all run in about 16 seconds, achieving close to 30 GB/s of bandwidth. Non-copy variants (Pochoir, hand-coded, and OPS) run in 8.3-9.4 seconds, with a bandwidth around 30 GB/s. This clearly shows that at this point performance is bound by how much data is moved, with a 2$\times$ performance difference between the copy and non-copy versions as expected, due to the difference in the amount of data moved. When cache-blocking tiling is enabled, performance improves dramatically, up to 4.5$\times$, and the difference between copy and non-copy versions is less than 1.5$\times$, showing that computations and latency of instructions starts to dominate performance. Copy versions of Pluto and the hand-coded test perform almost exactly the same (3.5 vs. 3.54 seconds), despite using completely different tiling and parallelisation approaches (diamond tiling with parallelisation over tiles vs. plain skewed tiling with parallelisation within tiles). Both achieve 140 GB/s overall, which is a very good fraction of peak L3 bandwidth, considering misaligned accesses due to stencils and the cache flushes between tiles. In contrast to all other versions, OPS constructs tiles based on run-time information only, and calls computational kernels through function pointers which adversely affects performance as predicted by the outlined hand-coded benchmark, nevertheless it still achieves a 3.1$\times$ speedup over the non-tiled version.

\subsection{Tiling CloverLeaf}

The previous benchmark showed the superiority of stencil/polyhedral compilers, however they cannot be applied to an application like CloverLeaf at sufficient scope: code analysis and experiments showed that only a handful of code blocks with 3-4 consecutive loopnests can be tiled across, because either the sequence of loops cannot be determined at compile time or subsequent loops are in different compilation units. Tiling them with Pluto did not result in a measurable performance difference.

In contrast, OPS determines the loops to tile across at runtime. As it does not require any modifications to user code, it is possible to automatically deploy this optimisations to large-scale codes.

Enabling automatic tiling in OPS requires using a different code generator at compile-time, but otherwise no action is necessary on the part of the user. At runtime, unless specified otherwise, OPS will tile over all loops up to the point where a global reduction is reached, to make sure control decisions are executed correctly in the host code. In CloverLeaf, this means tiling over an entire time iteration, which is a sequence of 153 loops in 2D and 603 loops in 3D - in this paper we do not parametrise the number of loops to tile over, as this would be highly non-trivial given the variety of loops (there are a lot of boundary condition loops with very thin iteration ranges), this will be addressed in future work.

Aside from the automatic tile size selection algorithm, it is also possible to manually specify tile sizes in each dimension. Since OPS is parallelising within each tile, we size the tiles in the last dimensions to be an integer multiple of the number of threads - in 2D this is achieved by setting the Y tile size to be a factor of 20, and in 3D we collapse the Y and Z loops (using OpenMP pragmas), and set Y and Z so that their product is a factor of 20.

Here, we study the performance of both the 2D and the 3D versions of CloverLeaf. In 2D, we use a $6144^2$ mesh, and run 10 time iterations, and in 3D, we use a $330^3$ mesh, and run 10 time iterations. The total memory footprint in 2D is 7.054 GB and in 3D 8.741 GB. Note that normally CloverLeaf would run for over 10000 time iterations to fully resolve the simulation, but since its execution is following a recurring pattern, here we can restrict it to 10 and still obtain representative performance figures.

\begin{table*}[t]\small
\centering
\caption{Peformance of CloverLeaf 2D baseline and tiled\normalsize}
\vspace{-6pt}\small
\begin{tabular}{ l | cccc | ccccc }
  \hline			
                  & \multicolumn{4}{c|}{CloverLeaf 2D OpenMP}& \multicolumn{5}{c}{CloverLeaf 2D Tiled} \\ \hline \hline
Phase             & Time(sec) & \% & GB/s & GFLOPS/s & Time(sec) & \% & GB/s & GFLOPS/s & Speedup \\ \hline
Timestep          &   0.71 &   3.79 & 39.67 & 58.57 & 0.69 & 7.33  & 40.90  & 60.40  & 1.03 \\ \hline
Ideal Gas         &   0.89 &   4.78 & 30.21 & 30.41 & 0.65 & 6.97  & 41.27  & 41.54  & 1.37 \\ \hline
Viscosity         &   0.89 &   4.75 & 15.86 & 58.76 & 0.36 & 3.86  & 38.85  & 143.92 & 2.45 \\ \hline
PdV               &   2.36 &  12.64 & 30.97 & 30.37 & 1.54 & 16.46 & 47.36  & 46.45  & 1.53 \\ \hline
Revert            &   0.37 &   1.97 & 30.63 & 0.00  & 0.08 & 0.83  & 143.86 & 0.00   & 4.70 \\ \hline
Acceleration      &   0.92 &   4.90 & 33.80 & 21.44 & 0.40 & 4.28  & 77.10  & 48.92  & 2.28 \\ \hline
Fluxes            &   0.62 &   3.32 & 36.24 & 7.30  & 0.34 & 3.67  & 65.37  & 13.16  & 1.80 \\ \hline
Cell Advection    &   4.01 &  21.46 & 30.16 & 18.92 & 2.23 & 23.73 & 54.34  & 34.09  & 1.80 \\ \hline
Momentum Advection&   6.68 &  35.77 & 32.84 & 12.89 & 1.95 & 20.77 & 112.66 & 44.21  & 3.43 \\ \hline
Reset             &   0.74 &   3.95 & 30.46 & 0.00  & 0.25 & 2.62  & 91.51  & 0.00   & 3.00 \\ \hline
Update Halo       &   0.07 &   0.39 &  2.66 & 0.00  & 0.26 & 2.80  & 0.73   & 0.00   & 0.28 \\ \hline
Field Summary     &   0.11 &   0.57 & 47.82 & 37.44 & 0.05 & 0.56  & 96.42  & 75.50  & 2.02 \\ \hline
The Rest          &   0.31 &   1.68 &  6.26 & 13.09 & 0.52 & 5.53  & 3.80   & 7.94   & 0.61 \\ \hline \hline
Total             &  18.68 & 100.00 & 30.89 & 20.71 & 8.73 & 100.0 & 66.12  & 44.31  & 2.14 \\
\end{tabular}\label{tab/cloverleaf_2d}
\end{table*}\normalsize

\subsubsection{Baseline performance}
The baseline performance using pure OpenMP restricted to a single socket is established at 18.7 seconds in 2D and 32.5 seconds in 3D. All innermost loops are reported as vectorised by the compiler. We show performance breakdowns in Tables \ref{tab/cloverleaf_2d} and \ref{tab/cloverleaf_3d} - note that here parallel loops are grouped by computational phase and averaged (weighted with relative execution time). It is clear that bandwidth is the key bottleneck in both 2D and 3D - especially in 2D where average bandwidth is 30 GB/s, same as on the Heat equation. There are a number of more computationally intensive kernels, where a considerable fraction of the peak computational throughput is achieved: for example Viscosity achieves 58/62 GFLOPS/s - but this is still not high enough to tip the balance from bandwidth-bound to compute-bound. Average bandwidth over the entire application in 3D is reduced to 25 GB/s, and the average computational throughput decreases from 20.7 to 18.9 GFLOPS/s. It is therefore clear that for most of these applications, the code is bound by bandwidth, which could potentially be improved with cache-blocking tiling.


\begin{table*}[t]\small
\centering
\caption{Performance of CloverLeaf 3D baseline and tiled\normalsize}
\vspace{-6pt}\small
\begin{tabular}{ l | cccc | ccccc }
  \hline      
                  & \multicolumn{4}{c|}{CloverLeaf 3D OpenMP}& \multicolumn{5}{c}{CloverLeaf 3D Tiled} \\ \hline \hline
Phase             & Time(sec) & \% & GB/s & GFLOPS/s & Time(sec) & \% & GB/s & GFLOPS/s &Speedup \\ \hline
Timestep          &    1.77 &    5.27 &  18.14 & 31.64 &  0.85 &    5.05 &  38.02 & 66.34  & 2.10 \\ \hline
Ideal Gas         &    0.89 &    2.64 &  28.99 & 29.18 &  0.64 &    3.80 &  40.36 & 40.63  & 1.39 \\ \hline
Viscosity         &    1.71 &    5.10 &  14.06 & 62.39 &  0.63 &    3.74 &  38.54 & 171.01 & 2.74 \\ \hline
PdV               &    3.88 &   11.54 &  21.41 & 24.84 &  2.47 &   14.77 &  33.56 & 38.94  & 1.57 \\ \hline
Revert            &    0.37 &    1.09 &  29.22 & 0.00  &  0.12 &    0.72 &  88.73 & 0.00   & 3.04 \\ \hline
Acceleration      &    2.05 &    6.11 &  18.42 & 19.43 &  1.08 &    6.47 &  34.92 & 36.83  & 1.90 \\ \hline
Fluxes            &    1.13 &    3.36 &  28.58 & 9.59  &  0.48 &    2.84 &  67.70 & 22.72  & 2.37 \\ \hline
Cell Advection    &    6.46 &   19.22 &  27.65 & 17.18 &  2.97 &   17.74 &  60.14 & 37.36  & 2.17 \\ \hline
Momentum Advection&   12.82 &   38.16 &  29.40 & 13.50 &  4.50 &   26.91 &  83.67 & 38.41  & 2.85 \\ \hline
Reset             &    0.91 &    2.71 &  29.60 & 0.00  &  0.68 &    4.06 &  39.63 & 0.00   & 1.34 \\ \hline
Update Halo       &    0.99 &    2.94 &  12.65 & 0.00  &  1.55 &    9.29 &  8.04  & 0.00   & 0.64 \\ \hline
Field Summary     &    0.17 &    0.50 &  33.45 & 45.77 &  0.07 &    0.44 &  75.53 & 103.34 & 2.26 \\ \hline
The Rest          &    0.45 &    1.34 &   5.56 & 13.92 &  0.52 &    3.12 &  4.79  & 11.99  & 0.86 \\ \hline \hline
Total             &   32.5  &  100.00 &  25.27 & 18.87 &  16.56&  100.00 &  51.24 & 38.26  & 1.96 \\
\end{tabular}\label{tab/cloverleaf_3d} 
\end{table*}\normalsize

\subsubsection{Tiling CloverLeaf 2D}
After enabling tiling in OPS, the best performance is achieved at a tile size of $640\times160$ - 8.73 seconds. At this point the memory footprint of the tile is 20.4 MB, and the speedup over the non-tiled version is 2.13$\times$. Tiled performance is very resilient to changes in the exact tile size: in our experiments there were 32 tile size combinations out of 144 within 2\% of the optimum performance, with up to 60\% smaller tiles, or up to 20\% bigger tiles. For more details, please see the Supplementary Material. The tile size automatically chosen by OPS for the main time iteration is $600\times200$, with a runtime of 8.82 seconds, 1\% slower that the tile size found by searching. The overhead of computing the tiling plans and loop schedules is 0.016 seconds, or 0.18\% of the total runtime, which would be further diminished on longer runs.

Performance breakdowns for the 2D version are shown in Table \ref{tab/cloverleaf_2d} - in line with the overall reduction in runtime, both average bandwidth and computational throughput have more than doubled. It also shows that the least computationally intensive loops have improved the most; most of these are straightforward loops with few stencils, such as Revert and Reset which copy from one dataset to another: here we see a 3-4.5$\times$ improvement in runtime compared to non-tiled versions. The single most expensive phase is Momentum advection, with several fairly simple kernels, it gains a 3.4$\times$ speedup. In contrast, loops in Timestep or PdV gain only 1.1-1.5x due to reductions and higher computational intensity. Notably, boundary loops in Update Halo slow down by a factor of up to 3.7 - this is due to small loops being subdivided even further, worsening their overheads. Viscosity now achieves 143 GFLOPS/s, a considerable fraction of the measured peak. The second most expensive computational phase is cell advection, it gains only a 1.8$\times$ speedup, which is in part due to the large number of branches within the computational kernels.


\subsubsection{Tiling CloverLeaf 3D}
For CloverLeaf 3D, the best performance is achieved when the X dimension is not tiled, and the Y and Z tile sizes are both 20 - 16.2 seconds, which is 2$\times$ faster than the baseline. Once again, we see the performance being resilient to the exact tile size, although the number of possible tile size combinations is significantly less than in 2D; out of 80 combinations tested, there were 6 other tile size combinations within 2\% of the best performance, and 18 within 10\%. The tile size automatically chosen by OPS is $330\times17\times17$, with a runtime of 17.02 seconds, or 5\% slower that the tile size found by search. The overhead of computing the tiling plans and loop schedules is 0.01 seconds.

Performance breakdowns in Table \ref{tab/cloverleaf_3d} show a more consistent speedup over the baseline version compared to the 2D version: in 2D the standard deviation of speedups is 1.2, whereas in 3D it is only 0.75. The slowdown on Update halo is only 1.5$\times$, and the best speedup (Revert) is only 3$\times$. Both overall bandwidth and computational throughput have increased by a factor of two - Viscosity now achieves 171 GFLOPS/s, or 63\% of peak, and the highest achieved bandwidth is 88 GB/s on Revert.

Enabling reporting from the dependency analysis in OPS shows the amount of skewing between the ``bottom'' and the ``top'' of the tiles; due to the large number of temporary datasets and the way loops are organised in directional sweeps, the total skewing is only 12 grid points in each direction in 2D and 14 grid points in the Y and Z directions in 3D (note that in this benchmark we do not tile in the X direction). In 2D, the additional data needed due to this skew is only a small fraction of the total tile size of $640\times160$ (0.2\%), leading to almost perfect data re-use within tiles. However, in 3D where the Y and Z tile sizes are $20$, a skew of 14 points is 49\%, meaning that much of the data is replaced between the execution of loops at the ``bottom'' of the tile and the ``top'' of the tile, nevertheless data re-use is still very high, the replacement happens gradually and it can be served fast enough from off-chip memory.

\subsection{Tiling TeaLeaf and OpenSBLI}

To demonstrate and underline the applicability of tiling in OPS to different applications as well, we deploy and evaluate this optimisations to two more applications, the TeaLeaf iterative solver code and the OpenSBLI compressible fluids solver. For TeaLeaf, we choose a $2000^2$ mesh, and the PPCG solver, configured so that most time is spent in the preconditioning phase which does not require reductions (which limits tiling and is generally a bottleneck over MPI particularly), for the full list of configuration parameters, see \cite{ops-github} and \texttt{tea\_tiling.in}. Keeping the benchmark still representative, we restrict the execution to 4 time iterations. The total memory footprint is 628 MB, however, the bulk of the runtime is spent executing the preconditioner, which accesses only 5 datasets - a total of 160 MB. For OpenSBLI, we use a $257^3$ grid and the RS variant \cite{opensbli}, solving the Taylor-Green vortex testcase, and limiting the number of time iterations to 10; running to completion - 500 iterations - follows the same execution patterns, thus our 10 iterations are representative. The total memory footprint is 3.84 GB.

The results are presented in Table \ref{tab/other}: runtime without tiling, with tiling and the automatic tile size selection algorithm and with using the best tile size found by exhaustive search ($1010\times400$). The performance gain on TeaLeaf by enabling tiling is up to 3.5$\times$, with a 5\% difference between the automatic and manual tile size selection. The large speedup on TeaLeaf is largely due to the bulk of the runtime being spent in just two key loops doing preconditioning, without the need for reductions in-between. There are just 5 datasets being accessed in that part of the algorithm, and the computations themselves are fairly simple - only multiplications and additions. The achieved bandwidth for the baseline version is 46.27 GB/s (95\% of peak), and the computational throughput is 13.3 GFLOPS/s. With tiling enabled, achieved bandwidth increases to 164.7 GB/s and computational throughput to 47.35 GFLOPS/s.

Due to the small number of datasets used in key parts of the algorithm, tiling performance in TeaLeaf is very sensitive to the problem size: if we reduce it to $700^2$ (20MB memory footprint), the speedup is reduced to just $1.07\times$. Performance is also very sensitive to the solver selection - if we use the Conjugate Gradient solver, without preconditioning, there is very little speedup at any problem size ($1.05\times$ at $4000^2$), due to the frequent reductions, which prohibit tiling across more than 2 loops.

Given that for most of the runtime only two key loops repeat for a large number of times for this testcase, we manually modified the code to enable tiling by Pluto - this involved multiversioning at a fairly high level in the call stack (something that we argue also makes the code more difficult to read and maintain). With the default tile size selection algorithm in Pluto, performance actually decreased by $1.9\times$ to 25.66 seconds. After an exhaustive search over potential tiles sizes (lasting for several hours) we have only been able to improve upon the non-tiled performance by $2.44\times$ to 5.49 seconds (tile size of $192\times20$ and 5 tile height).

Performance results from OpenSBLI are also presented in Table \ref{tab/other}: here speedup from tiling is lower than on other applications: $1.7\times$. The performance difference between automatic and manual tile selection here is 9\%. In this application, 60\% of the runtime is spent in an extremely complicated computational loop containing arithmetic expressions that are tens of lines long (5 expressions in 151 lines), uses a large number of sqrt operations, and accesses 167 double precision values per grid point. As such it is not limited by either bandwidth (7.16 GB/s) or computational throughput (27.7 GFLOPS/s), rather by latency, register pressure, and other factors. Nevertheless, it still gains a $1.57\times$ speedup from tiling, achieving 11.24 GB/s and 43.49 GFLOPS/s. Overall, the entire application gains a $1.71\times$ improvement from tiling.

\begin{table}\small
\centering
\caption{Performance summary of TeaLeaf and OpenSBLI\normalsize}
\vspace{-6pt}\small
\begin{tabular}{ l | c | c | c | r }
  \hline			
  Test & Baseline & Tiled & Tiled Best & Speedup \\ \hline \hline
  TeaLeaf &  13.438s & 3.9339s & 3.7666s & 3.56$\times$ \\ \hline
  OpenSBLI & 20.807s & 13.221s & 12.1385s & 1.71$\times$ \\ \hline
\end{tabular}\label{tab/other}
\end{table}\normalsize

\section{Tiling in distributed-memory systems}

To evaluate the efficiency of our tiling approach, we deploy the codebase on CINECA's Marconi supercomputer, and benchmark both strong scaling and weak scaling of the four large-scale codes. Marconi's A1 phase consists of 1512 nodes, each with dual-socket 18-core Broadwell Xeon E5-2697 v4 CPUs, running at 2.3 GHz (Hyper-Threading is off). The nodes are interconnected with Intel's 100Gb/s OmniPath fabric. The scheduling system currently limits job sizes to 160 nodes, therefore in our power-of-two scaling studies, we evaluate performance on up to 128 nodes, or 4608 cores. 

For strong scaling CloverLeaf, we use the same mesh sizes as on the single socket tests. For strong scaling TeaLeaf and OpenSBLI we take the same problem sizes as described previously, and double their size in the x direction (due to the particularly small problem sizes in TeaLeaf). To evaluate scalability, we then keep this problem size and run it on an increasing number of nodes. Thus for CloverLeaf 2D, we strong scale a $6144^2$ problem for 10 time iterations, for CloverLeaf 3D, a $330^3$ problem for 10 time iterations, for TeaLeaf, a $4000\times2000$ problem for 2 solver iterations, and for OpenSBLI, a $514\times257\times257$ problem, for 10 time iterations.

For weak scaling, we take the problem sizes described for strong scaling, then scale it with the number of nodes, keeping the per-node size constant. For TeaLeaf, due to its convergence-dependent control flow that changes as we increase the problem size, we also alter convergence criteria to keep the number of solver iterations and the preconditioning iterations approximately the same - as it is not feasible to control this so that the number of iterations matches exactly, we report performance as time per 100 preconditioning iterations.

\begin{figure}[t!]
\hspace*{-10pt}\centering
\includegraphics[width=0.52\textwidth]{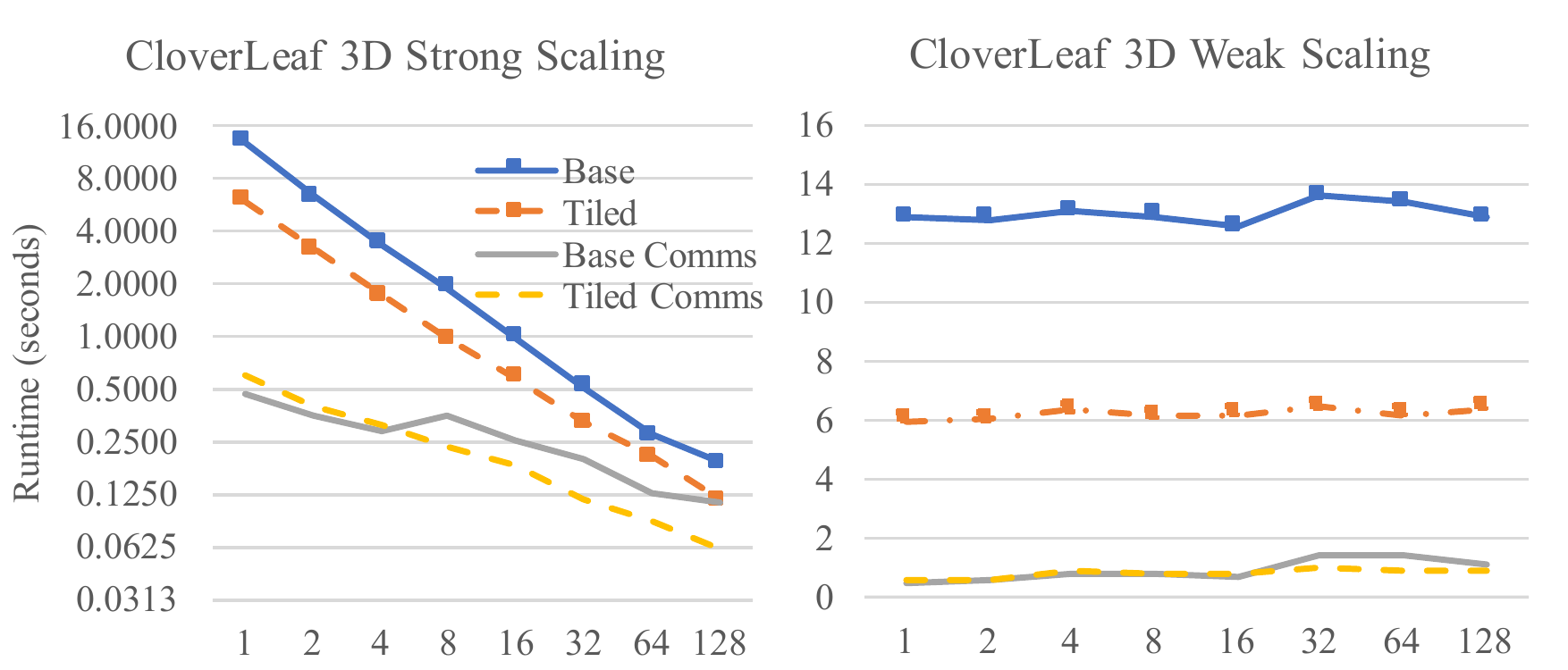}\vspace{-12pt}
\caption{\small Scaling CloverLeaf 3D to multiple nodes on Marconi\normalsize}\vspace{-10pt}\label{fig/cloverleaf_mpi_3d}
\end{figure}

\begin{figure}[t!]
\hspace*{-10pt}\centering
\includegraphics[width=0.52\textwidth]{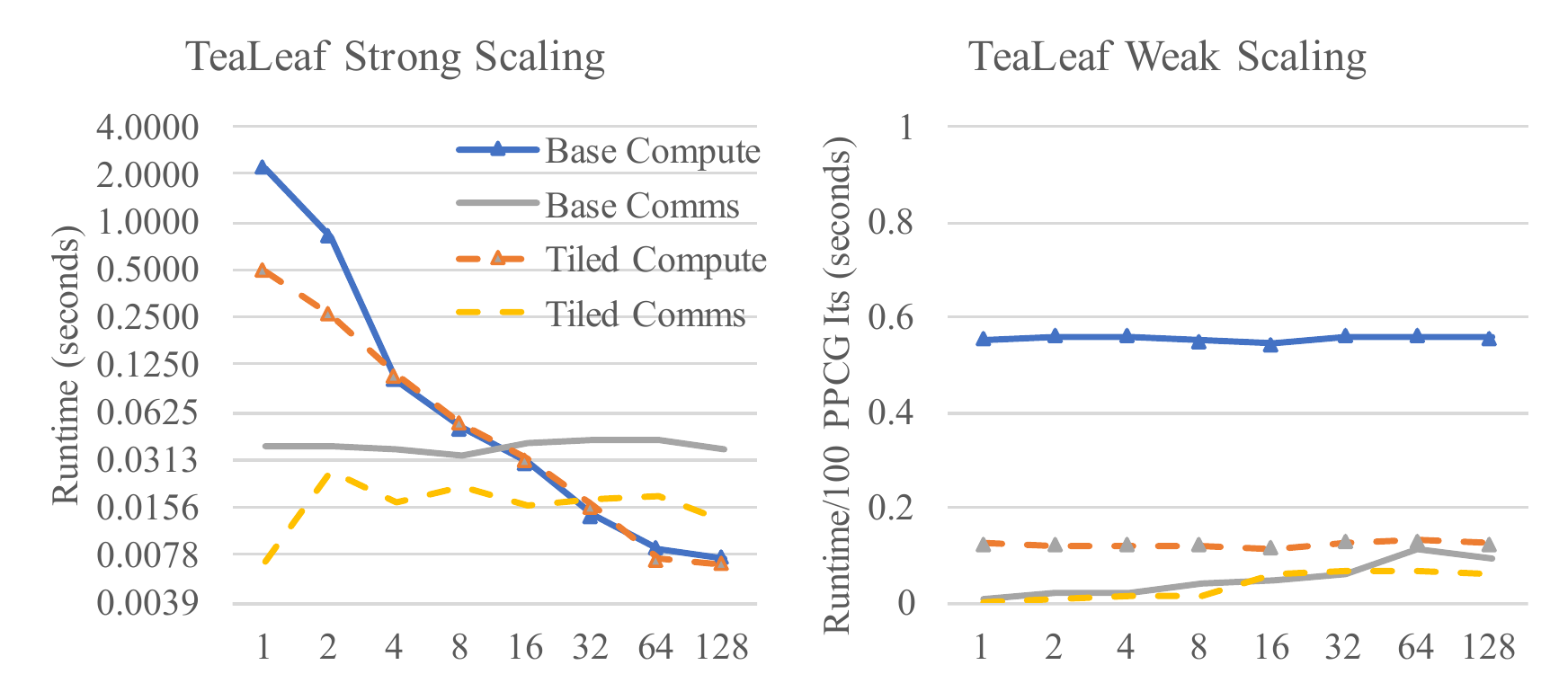}\vspace{-12pt}
\caption{\small Scaling TeaLeaf to multiple nodes on Marconi\normalsize}\vspace{-10pt}\label{fig/tealeaf_mpi}
\end{figure}

Figures \ref{fig/cloverleaf_mpi_3d}-\ref{fig/tealeaf_mpi} show the results of these scaling tests for CloverLeaf 3D and TeaLeaf - CloverLeaf 2D and OpenSBLI look very similar to CloverLeaf 3D, they are deferred to the Supplementary Material. For CloverLeaf 3D, we show total runtime as well as the time spent in MPI communications. For TeaLeaf, we show the time spent computing and the time spent in MPI communications separately (total time not shown), due to the much larger relative cost of communications. Tiling over MPI scales well, keeping the speedup ratio between tiled and non-tiled versions at higher node counts as well, except for the strong scaled TeaLeaf test, where after 4 nodes the two computational times are nearly identical: a quick calculation shows that at 4 nodes (8 sockets) the preconditioner's memory footprint is 40MB (versus the 35 MB L3 cache size), therefore even without tiling, the datasets stay in cache. Time spent in MPI communications, particularly when strong scaling,  shows the advantage of the communications scheme used when tiling, which effectively results in fewer but larger messages. 

\section{Tiling on Intel Knights Landing}

The second-generation Intel Xeon Phi platform, also called Knights Landing (KNL) has a 16GB on-chip stacked memory, with a $4-5\times$ higher bandwidth than that of DDR4. This memory can serve either as a cache to off-chip DDR4 memory, or as a separately managed memory space, or a combination of the two with a pre-defined split. This high bandwidth stacked memory is a great benefit to applications bound by memory bandwidth, such as our stencil codes. However, if the problem size has a larger memory footprint than 16GB, the user either has to manually allocate different datasets to different memory spaces, or has to rely on good enough caching behaviour. Here, we study the latter case and demonstrate how our tiling approach can maintain high performance and cache efficiency even when the full problem size is much larger than 16GB. We did not experience any benefit from trying to size tiles so they stay in L2 cache, therefore we do not report on those experiments. 
\begin{figure}[t!]
\hspace*{-10pt}\centering
\includegraphics[width=0.40\textwidth]{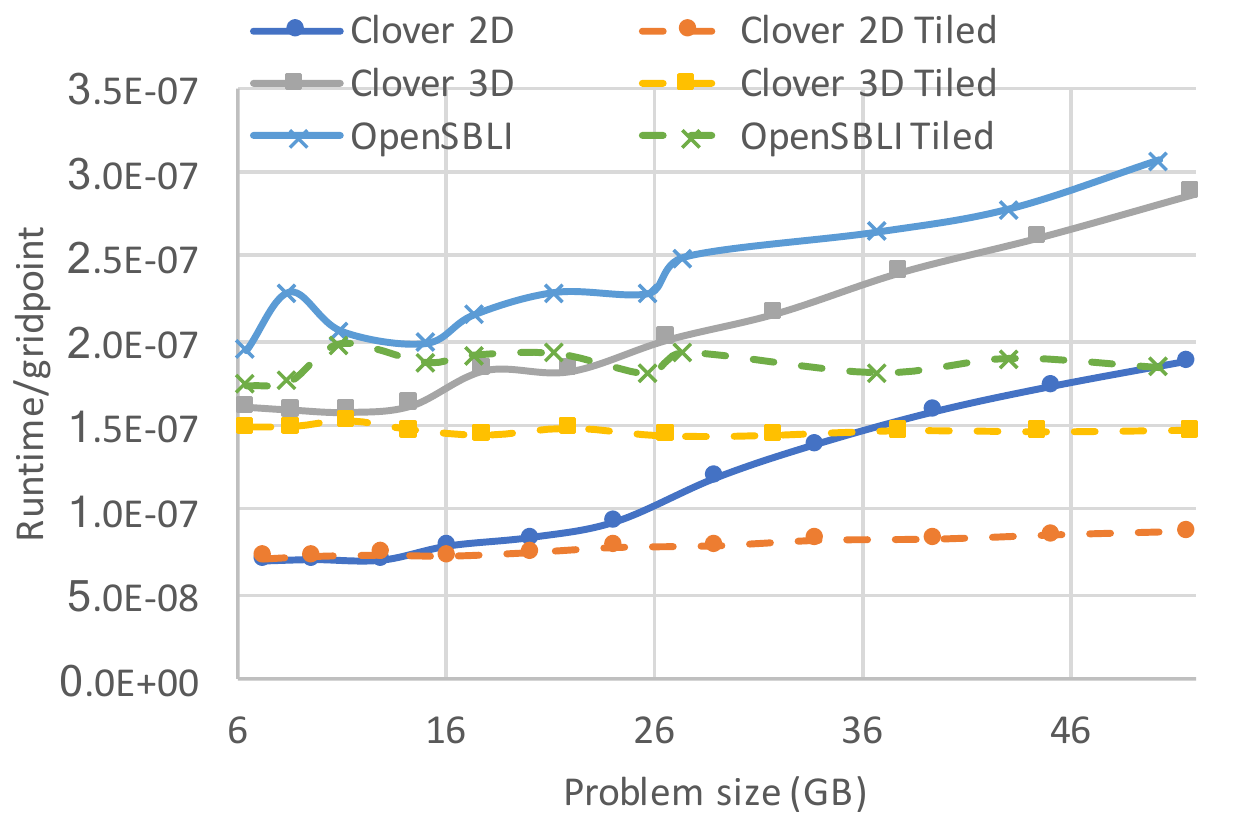}\vspace{-12pt}
\caption{\small Problem size scaling on KNL\normalsize}\vspace{-10pt}\label{fig/knl}
\end{figure}

Figure \ref{fig/knl} shows the performance of the tiled and non-tiled implementations of CloverLeaf 2D/3D and OpenSBLI when the problem size increases; here we normalised runtimes by dividing them with the number of gridpoints. On our x200 7210 chip (64 cores), we run with 4 MPI processes and 64 OpenMP threads each, and a cache/quadrant memory configuration. As the figure shows, up to a size of 16GB, all data fits in cache, and there is no or very little benefit (due to improved communications) from tiling, and the runtime per gridpoint remains constant. Beyond 16GB however, the non-tiled versions gradually slow down as less and less of the data being worked on stays resident in cache. Tiled versions on the other hand maintain their runtime per gridpoint, demonstrating the utility of our tiling approach even on a machine with orders of magnitude larger cache. At the larger problem sizes where the memory footprint is approximately 50GB, there is a $2.17\times$ speedup on CloverLeaf 2D, a $1.95\times$ speedup on CloverLeaf 3D, and a $1.67\times$ speedup on OpenSBLI. TeaLeaf was omitted due to its small memory footprint at reasonable problem sizes.

\begin{figure}[t!]
\hspace*{-10pt}\centering
\includegraphics[width=0.54\textwidth]{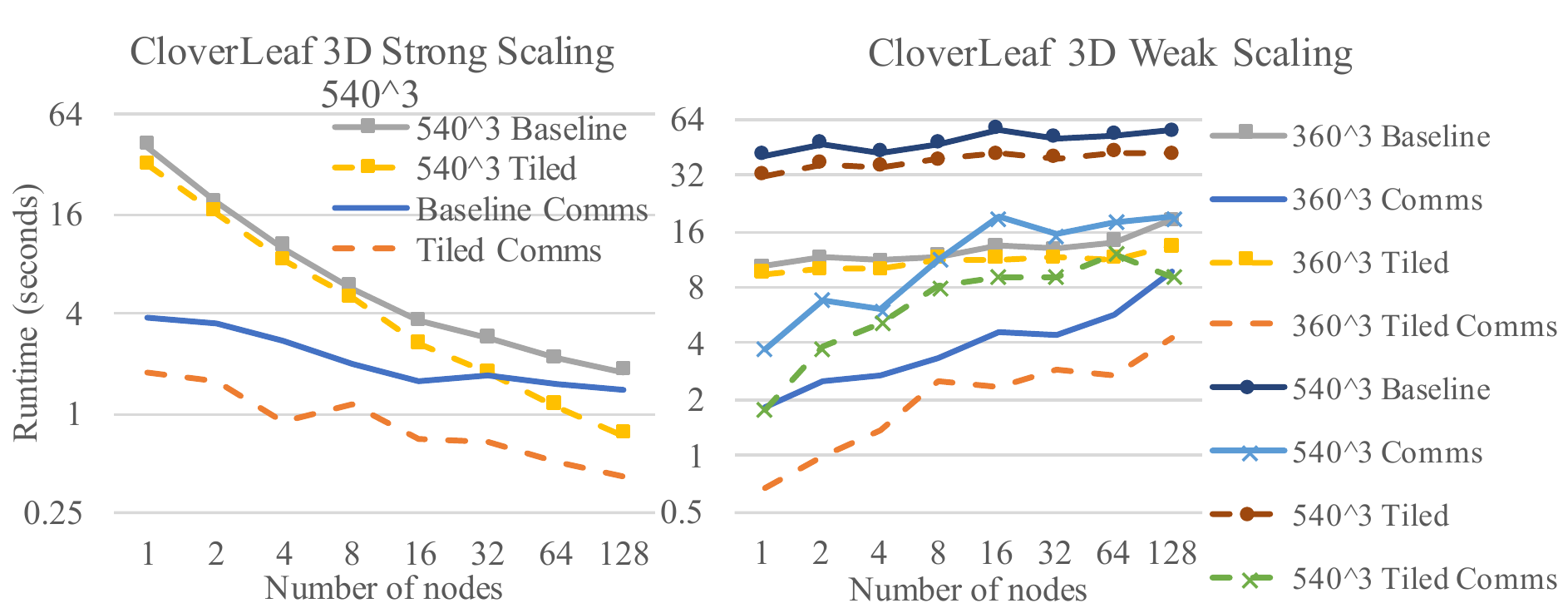}\vspace{-12pt}
\caption{\small Scaling CloverLeaf 3D on Marconi-A2 (KNL)\normalsize}\vspace{-10pt}\label{fig/knl_mpi}
\end{figure}

We also evaluate scaling up to 128 nodes on Marconi-A2, which has Intel Xeon Phi x200 7250 nodes (68 cores each). We weak scale two cases; one where all data fits in the 16GB cache and one where it does not. Here we report on CloverLeaf 3D, where we scaled a $360^3$ (11.1 GB) and a $540^3$ (37.8 GB) mesh - results are shown in Figure \ref{fig/knl_mpi}. For both strong scaling and weak scaling the smaller mesh, the baseline and the tiled versions perform very similar since all data can stay resident in cache. The tiled version performs slightly better due to its improved communications scheme; the cost of MPI communications is on average $2\times$ less. When weak scaling the larger mesh though, we can consistently see a 25\% improvement over the baseline version. Scaling on CloverLeaf 2D and OpenSBLI show the same behaviour, their performance figures can be found in the Supplementary Material.

\section{Conclusions}
\label{sec/conc}

In this paper, we explored the challenges in achieving cache-blocking tiled execution on large PDE codes implemented using the OPS library; codes that are significantly larger than the traditional benchmarks studied in the literature: execution spans several compilation units, and the order of loops can not be determined at compilation time. The key issues included how to handle dynamic execution paths within and across loop nests, and across a number of source files - something state-of-the-art polyhedral and stencil compilers (Pluto, Pochoir) cannot do. To tackle this, we adopt the locality improving optimisations called iteration space slicing, and instead of trying to tile at compile-time, we develop a run-time capability that relies on building a chain of loops through a delayed execution scheme. 


To study the proposed approach, we established a baseline with a comparative study of the finite-difference heat equation Jacobi solver, comparing against Pluto and Pochoir. These libraries use different tiling and parallelisation strategies, and rely on compile-time analysis of the stencil code, but performance is closer than expected - most of the overhead is due to going through function pointers at run-time. Overall, the OPS version achieves a 3.1$\times$ speedup over the non-tiled version and a bandwidth of 97.8 GB/s.

Thanks to the run-time analysis, the proposed approach can be trivially applied to larger-scale applications as well; we study the 2D and 3D versions of the CloverLeaf application in detail. Establishing a baseline shows that both versions are bound by bandwidth to off-chip memory. Enabling tiling  shows a speedup of up to 2.1$\times$ in 2D and 2$\times$ in 3D. Detailed performance analysis shows that the simplest loops gain the most performance improvement, some computationally intensive loops become limited by compute instead of bandwidth, and that thin boundary loops slow down. We demonstrate that out results are immediately applicable to applications that use OPS, including TeaLeaf (achieving up to $3.56\times$) and OpenSBLI (achieving up to $1.71\times$).

Our algorithms and testing are also extended to distributed memory systems, demonstrating excellent scalability, maintaining speedup over the non-tiled versions as long as the problem size per socket is reasonably larger than the cache size. Performance is further improved by the communications scheme, particularly when strong scaling, due to using fewer but larger messages. Our work is also evaluated on Intel's Knights Landing platform, showing that even on an architecture with a much larger cache (16GB) our algorithms provide significant performance improvements when the full problem size grows beyond the capacity of the cache, and this improvement is maintained when weak scaled up to 128 nodes.

The fact that cache-blocking tiling can be applied with such ease to larger, non-trivial applications once again underlines the utility of domain specific languages, and their main premise: once an application is implemented using a high-level abstraction, it is possible to transform the code to near-optimal implementations for a variety of target architectures and programming models, without any modifications to the original source code. The algorithms presented in this paper are generally applicable to any stencil DSL that provides per loop data access information.

Future work includes the study of tile height; the number of loops to be tiled over, which is highly non-trivial in the presence of such a diverse set of loops.

\section*{Acknowledgments}

The authors would like to thank Michelle Strout at the University of Arizona for her invaluable suggestions and insight, as well as Fabio Luporini at Imperial College London. This paper was supported by the J\'anos B\'olyai Research Scholarship of the Hungarian Academy of Sciences.

The OPS project is funded by the UK Engineering and Physical Sciences Research Council projects EP/K038494/1,
EP/K038486/1, EP/K038451/1 and EP/K038567/1 on ``Future-proof massively-parallel execution of multi-block applications'' project. This research was also funded by the Hungarian Human Resources Development Operational Programme (EFOP-3.6.2-16-2017-00013).
We acknowledge PRACE for awarding us access to resource Marconi based in Italy at Cineca. 

\vspace{-7pt}

\bibliographystyle{IEEEtran}
\vspace{-10pt}
\bibliography{tiling}

\vspace{-40pt}
\begin{IEEEbiography}[{\includegraphics[width=1in,height=1.25in,clip,keepaspectratio]{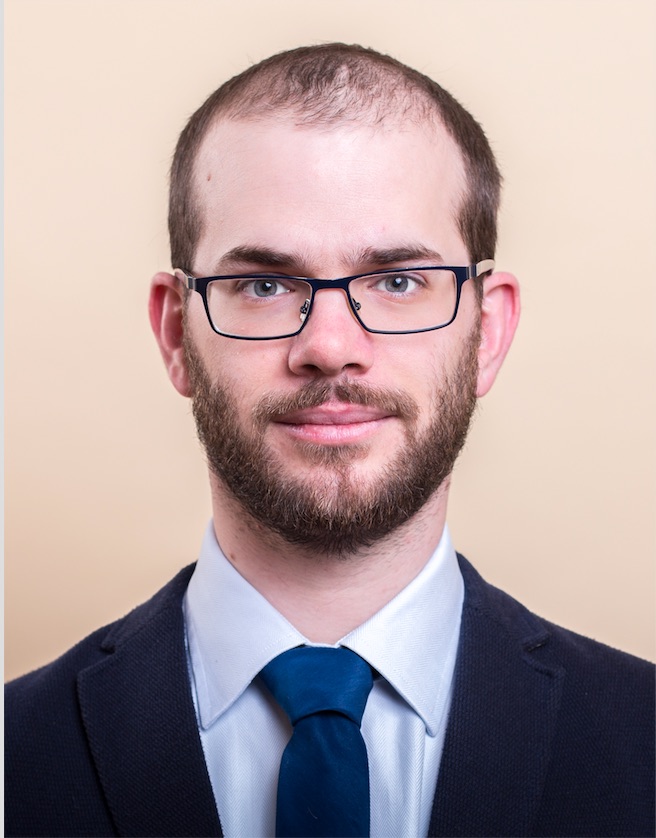}}]{Istv\'an Z. Reguly} is a lecturer at PPCU ITK, Hungary. He holds an MSc and a PhD in computer science from the PPCU, Hungary. His research interests include high performance scientific computing on many-core hardware and domain specific active libraries for structured and 
unstructured meshes.
\end{IEEEbiography}
\vspace{-48pt}	
\begin{IEEEbiography}[{\includegraphics[width=1in,height=1.25in,clip,keepaspectratio]{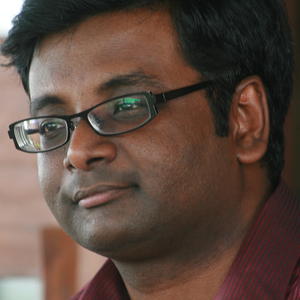}}]{Gihan R. Mudalige}
is an assistant professor at the Department of Computer Science, University of Warwick, UK. His research interests are in 
performance analysis/optimization of scientific applications on parallel, high-performance systems. Previously he has 
worked as a research fellow at the High Performance Systems Group at the University of Warwick and a research intern 
at the University of Wisconsin–Madison, US. Dr. Mudalige holds a PhD. in Computer Science from the University of Warwick 
and is a member of the ACM.
\end{IEEEbiography}
\vspace{-35pt}	
\begin{IEEEbiography}[{\includegraphics[width=1in,height=1.2in,clip,keepaspectratio]{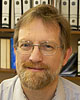}}]{Michael B. Giles}	
is Professor of Scientific Computing in Oxford University's Mathematical Institute where he carries out research 
into the development and analysis of more efficient Monte Carlo methods for computational finance and engineering 
uncertainty quantification.  He leads research into the use of GPUs for a variety of applications.
\end{IEEEbiography}
\end{document}